\documentclass{amsart}

\def\fH{{\mathcal H}} 
\def\fL{{\mathcal L}} 
\def\fS{{\mathfrak S}} 
\newtheorem{theorem}{Theorem}[section]

\theoremstyle{definition}
\newtheorem{definition}[theorem]{Definition}

\newtheorem{Con}[theorem]{Conjecture}

\newtheorem{Cor}[theorem]{Corollary}

\theoremstyle{remark}
\newtheorem{remark}[theorem]{Remark}

\numberwithin{equation}{section}

\begin{document}

\title[Trace inequalities on a generalized... ]
{Trace inequalities on a generalized Wigner-Yanase skew information}

\author{S. Furuichi}
\address{Department of Computer Science and System Analysis, College of Humanities and Sciences, Nihon University,3-25-40, Sakurajyousui, Setagaya-ku, Tokyo, 156-8550, Japan}
\email{furuichi@cssa.chs.nihon-u.ac.jp}
\thanks{The first author (S.F.) was supported in part by the Japanese Ministry of Education, Science, Sports and Culture, Grant-in-Aid for 
Encouragement of Young Scientists (B), 20740067 and Grant-in-Aid for Scientific Research (B), 18300003.}

\author{K.Yanagi}
\address{Division of Applied Mathematical Science, Graduate School of Science and Engineering, Yamaguchi University, Tokiwadai 2-16-1, Ube City, 755-0811, Japan}
\email{yanagi@yamaguchi-u.ac.jp}
\thanks{The second author (K.Y.) was also partially supported by the Ministry of Education, Science, Sports and Culture, Grant-in-Aid for Scientific Research (B), 18300003.}

\author{K.Kuriyama}
\address{Division of Applied Mathematical Science, Graduate School of Science and Engineering, Yamaguchi University, Tokiwadai 2-16-1, Ube City, 755-0811, Japan}
\email{kuriyama@yamaguchi-u.ac.jp}

\subjclass[2000]{Primary 47A63; Secondary 94A17}

\date{}


\keywords{Trace inequality, Wigner-Yanase skew information, Wigner-Yanase-Dyson skew information and uncertainty relation}

\begin{abstract}
We introduce a generalized Wigner-Yanase skew information and then derive the trace inequality related to the uncertainty relation.
This inequality is a non-trivial generalization of the uncertainty relation derived by S.Luo for the quantum uncertainty quantity excluding the classical mixure.
In addition, several trace inequalities on our generalized Wigner-Yanase skew information are argued.
\end{abstract}

\maketitle


\section{Introduction}
 Wigner-Yanase skew information
\begin{eqnarray} 
I_{\rho}(H) &\equiv& \frac{1}{2} Tr\left[ \left(i\left[\rho^{1/2},H\right]\right)^2\right] \label{W-Y_skew}\\ 
&=& Tr[\rho H^2] -Tr[\rho^{1/2}H\rho^{1/2}H]  \nonumber 
\end{eqnarray}
was defined in \cite{WY}. 
This quantity can be considered as a kind of 
the degree for non-commutativity between a quantum state $\rho$ and an observable $H$.
Here we denote the commutator by $[X,Y] \equiv XY-YX$.
This quantity was generalized by Dyson
\begin{eqnarray*} 
I_{\rho,\alpha}(H) &\equiv& \frac{1}{2} Tr\left[ \left(i\left[\rho^{\alpha},H\right]\right)\left(i[\rho^{1-\alpha},H]\right)\right]   \\
&=&Tr[\rho H^2] -Tr[\rho^{\alpha}H\rho^{1-\alpha}H],\,\,\, \alpha \in [0,1]  
\end{eqnarray*}
which is known as the Wigner-Yanase-Dyson skew information. It is famous that the convexity of $I_{\rho,\alpha}(H)$ with respect to $\rho$
was successfully proven by E.H.Lieb in \cite{Lie}. 
From the physical point of view, an observable $H$ is generally considered to be an unbounded operator, however in the present paper, unless otherwise stated,
we consider $H \in B(\fH)$, where $B(\fH)$ represents the set of all bounded linear operators on the Hilbert space $\fH$,
as a mathematical interest. We also denote the set of all self-adjoint operators (observables) by $\fL_h(\fH)$ and the set of all density operators (quantum states)
by $\fS(\fH)$ on the Hilbet space $\fH$.
The relation between the Wigner-Yanase skew information and the uncertainty relation was studied in \cite{Luo0}.
Moreover the relation between the Wigner-Yanase-Dyson skew information and the uncertainty relation was studied in \cite{Kos,YFK}.
In our previous paper \cite{YFK}, we defined a generalized skew information and then derived a kind of an uncertainty relation. 
In the section 2, we introduce a new generalized Wigner-Yanase skew information.
On a generalization of the original Wigner-Yanase skew information, our generalization is different from 
the Wigner-Yanase-Dyson skew information and a generalized skew information defined in our previous paper \cite{YFK}.
Moreover we define a new quantity by our generalized Wigner-Yanase skew information and then we
derive the trace inequality expressing a kind of the uncertainty relation. 

\section{Trace inequalities on a generalized Wigner-Yanase skew information}
Firstly we review the relation between the Wigner-Yanase skew information and the uncertainty relation.
In quantum mehcanical system, the expectation value of an observable $H$ in a quantum state $\rho$ is expressed by $Tr[\rho H]$.
It is natural that the variance for a quantum state $\rho$ and an observable $H$ is defined by 
$V_{\rho}(H)\equiv Tr[\rho \left(H-Tr[\rho H]I\right)^2] = Tr[\rho H^2] -Tr[\rho H]^2$.
It is famous that we have the Heisenberg's uncerainty relation:
\begin{equation}   \label{HUL}
V_{\rho}(A) V_{\rho}(B) \geq \frac{1}{4}\vert Tr[\rho[A,B]]\vert^2 
\end{equation}
for a quantum state $\rho$ and two observables $A$ and $B$.
The further strong result was given by Schr\"odinger
$$V_{\rho}(A) V_{\rho}(B)-\vert Cov_{\rho}(A,B)\vert^2 \geq \frac{1}{4}\vert Tr[\rho[A,B]]\vert^2,  $$
where the covariance is defined by $Cov_{\rho}(A,B) \equiv Tr[\rho \left(A-Tr[\rho A]I\right)\left(B-Tr[\rho B]I\right)].$
However, the uncertainty relation for the Wigner-Yanase skew information failed. (See \cite{Luo0,Kos,YFK}.)
$$I_{\rho}(A)I_{\rho}(B) \geq \frac{1}{4}\vert Tr[\rho[A,B]]\vert^2.$$
Recently, S.Luo introduced the quantity $U_{\rho}(H)$ representing a quantum uncertainty excluding the classical mixture:
\begin{equation}  \label{cla_mix}
U_{\rho}(H)  \equiv \sqrt{V_{\rho}(H)^2 -\left( V_{\rho}(H)-I_{\rho}(H)\right)^2},
\end{equation}
then he derived the uncertainty relation on $U_{\rho}(H)$ in \cite{Luo1}:
\begin{equation}   \label{UL_U}
U_{\rho}(A)U_{\rho}(B) \geq  \frac{1}{4}\vert Tr[\rho[A,B]]\vert^2.
\end{equation}
Note that we have the following relation
\begin{equation}  \label{note1}
0 \leq I_{\rho}(H)\leq U_{\rho}(H)\leq V_{\rho}(H).
\end{equation}
The inequality (\ref{UL_U}) is a refinement of the inequality (\ref{HUL}) in the sense of (\ref{note1}).

In this section, we study one-parameter extended inequality for the inequality (\ref{UL_U}).

\begin{definition}
For $0\leq \alpha \leq 1$, a quantum state $\rho$ and an observable $H$, we define the Wigner-Yanase-Dyson skew information 
\begin{equation} \label{W-Y-D_skew}
I_{\rho ,\alpha } \left( H \right) \equiv \frac{1}{2}Tr\left[ \left(i\left[ {\rho ^\alpha  ,H_0 } \right]\right)\left(i\left[ {\rho ^{1-\alpha}  ,H_0 } \right]\right)   \right]
\end{equation}
and we also define
\[
J_{\rho ,\alpha } \left( H \right) \equiv \frac{1}{2}Tr\left[ \left\{ {\rho ^\alpha  ,H_0 } \right\} \left\{ {\rho ^{1-\alpha}  ,H_0 } \right\}      \right],
\]
where $H_0 \equiv H-Tr[\rho H]I$ and we denote the anti-commutator by $\left\{X,Y\right\}=XY+YX$.
\end{definition}
Note that we have
$$
\frac{1}{2}Tr\left[ \left(i\left[ {\rho ^\alpha  ,H_0 } \right]\right)\left(i\left[ {\rho ^{1-\alpha}  ,H_0 } \right]\right)   \right]
=\frac{1}{2}Tr\left[ \left(i\left[ {\rho ^\alpha  ,H } \right]\right)\left(i\left[ {\rho ^{1-\alpha}  ,H } \right]\right)   \right]
$$
but we have
$$
\frac{1}{2}Tr\left[ \left\{ {\rho ^\alpha  ,H_0 } \right\} \left\{ {\rho ^{1-\alpha}  ,H_0 } \right\}      \right]
\neq 
\frac{1}{2}Tr\left[ \left\{ {\rho ^\alpha  ,H } \right\} \left\{ {\rho ^{1-\alpha}  ,H } \right\}      \right].
$$
Then we have the following inequalities:
\begin{equation}\label{ineq_i} 
  I_{\rho,\alpha}(H) \leq  I_{\rho}(H) \leq J_{\rho}(H) \leq J_{\rho,\alpha}(H),
\end{equation}
since we have $Tr[\rho^{1/2}H\rho^{1/2}H] \leq Tr[\rho^{\alpha}H\rho^{1-\alpha}H]$. 
(See \cite{Bou,Fuj} for example.)
If we define 
\begin{equation}   \label{gen_U}
U_{\rho,\alpha}(H)\equiv \sqrt{V_{\rho}(H)^2 -\left( V_{\rho}(H)-I_{\rho,\alpha}(H)\right)^2},
\end{equation}
as a direct generalization of Eq.(\ref{cla_mix}),
then we have 
\begin{equation} \label{note2}
0\leq I_{\rho,\alpha}(H) \leq U_{\rho,\alpha}(H) \leq U_{\rho}(H) 
\end{equation}
due to the first inequality of (\ref{ineq_i}). We also have 
\begin{equation} \label{u_expression1}
U_{\rho,\alpha}(H) = \sqrt{I_{\rho,\alpha}(H) J_{\rho,\alpha}(H)}.
\end{equation}

\begin{remark}
From the inequalities (\ref{note1}), (\ref{ineq_i}) and (\ref{note2}), our situation is that we have
$$
0 \leq I_{\rho,\alpha}(H)\leq I_{\rho}(H) \leq U_{\rho}(H)
$$
and
$$
0\leq I_{\rho,\alpha}(H) \leq U_{\rho,\alpha}(H) \leq U_{\rho}(H).
$$ 
Therefore our first concern is the ordering between  $I_{\rho}(H)$ and $U_{\rho,\alpha}(H)$.
However we have no ordering between them. Because we have the following examples.
We set the density matrix $\rho$ and the observable $H$ such as
 \[
\rho  = \left( \begin{array}{l}
 \,0.6 \,\,\,\,\,\,\,0.48 \\ 
 0.48\,\,\,\,\,\,\,0.4 \\ 
 \end{array} \right),H = \left( \begin{array}{l}
 1.0\,\,\,\,\,0.5 \\ 
 0.5\,\,\,\,\,5.0 \\ 
 \end{array} \right).
\]
If $\alpha=0.1$, then $U_{\rho,\alpha}(H) -  I_{\rho}(H)$ approximately takes $-0.14736$.
If $\alpha=0.2$, then $U_{\rho,\alpha}(H) -  I_{\rho}(H)$ approximately takes $0.4451$.

\end{remark}

\begin{Con}
Our second concern is to show an uncertainty relation with respect to $U_{\rho,\alpha}(H)$ as a direct generalization of the inequality (\ref{UL_U}) such that
\begin{equation}  \label{conj_01}
U_{\rho,\alpha}(X)U_{\rho,\alpha}(Y)\geq \frac{1}{4}\vert Tr\left[\rho[X,Y]\right] \vert^2
\end{equation}
However we have not found the proof of the above inequality  (\ref{conj_01}). 
In addition, we have not found any counter-examples of the inequality (\ref{conj_01}) yet.
\end{Con}

In the present paper, we introduce a generalized Wigner-Yanase skew information which is a generalization of the Wigner-Yanase skew information defined in Eq.(\ref{W-Y_skew}),
but different from the Wigner-Yanase-Dyson skew information defined in Eq.(\ref{W-Y-D_skew}).

\begin{definition}
For $0\leq \alpha \leq 1$, a quantum state $\rho$ and an observable $H$, we define a generalized Wigner-Yanase skew information by
$$
K_{\rho,\alpha}(H) \equiv \frac{1}{2} Tr\left[ \left(i \left[ \frac{\rho^{\alpha}+\rho^{1-\alpha}}{2},H_0 \right]\right)^2\right]
$$
and we also define
$$
L_{\rho,\alpha}(H) \equiv \frac{1}{2} Tr\left[ \left( \left\{ \frac{\rho^{\alpha}+\rho^{1-\alpha}}{2},H_0 \right\}\right)^2\right].
$$
\end{definition}

\begin{remark}
For two generalized Wigner-Yanase skew informations $I_{\rho,\alpha}(H) $ and $K_{\rho,\alpha}(H) $, we have the relation:
$$
I_{\rho,\alpha}(H) \leq K_{\rho,\alpha}(H). 
$$
Indeed, for a spertral decomposition of $\rho$ such as $\rho = \sum_k \lambda_k \vert \phi_k \rangle \langle \phi_k \vert$,
we have the following expressions:
$$
I_{\rho,\alpha}(H) =
 \frac{1}{2} \sum_{m,n}  \left( \lambda_m^{\alpha} -\lambda_n^{\alpha} \right)\left(\lambda_m^{1-\alpha} -\lambda_n^{1-\alpha} \right) 
\vert \langle \phi_m\vert H \vert \phi_n \rangle \vert ^2
$$
and
$$
K_{\rho,\alpha}(H) =
 \frac{1}{2} \sum_{m,n} \left( \frac{\lambda_m^{\alpha} -\lambda_n^{\alpha} +\lambda_m^{1-\alpha} -\lambda_n^{1-\alpha}}{2} \right)^2
\vert \langle \phi_m\vert H \vert \phi_n \rangle \vert ^2.
$$ 
By simple calculations, we see
$$ 
\left( \frac{\lambda_m^{\alpha} -\lambda_n^{\alpha} +\lambda_m^{1-\alpha} -\lambda_n^{1-\alpha}}{2} \right)^2
- \left( \lambda_m^{\alpha} -\lambda_n^{\alpha} \right)\left(\lambda_m^{1-\alpha} -\lambda_n^{1-\alpha} \right) 
\geq 0.
$$
\end{remark}

Throughout this section, we put
$X_0 \equiv X - Tr[\rho X]I$ and $Y_0 \equiv Y - Tr[\rho Y]I$.
Then we show the following trace inequality.

\begin{theorem}\label{the2_1}
For a quantum state $\rho$ and observables $X,Y$ and $\alpha\in[0,1]$, we have
\begin{equation}
W_{\rho,\alpha}  \left( X \right) W_{\rho,\alpha}  \left( Y \right)
\geq 
\frac{1}{4}\left| {Tr\left[ {\left( {\frac{{\rho ^\alpha   + \rho ^{1 - \alpha } }}{2}} \right)^2 \left[ {X,Y} \right]} \right]} \right|^2 \label{ineq_the2_1}
\end{equation}
where
\[
W_{\rho,\alpha}  \left( X \right) \equiv 
\sqrt{   K_{\rho,\alpha}(X) L_{\rho,\alpha}(X) }.
\]
\end{theorem}
{\it Proof}:
Putting 
\begin{equation} \label{def_K}
M \equiv i \left[ \frac{\rho^{\alpha} +\rho^{1-\alpha}}{2},X_0  \right] x 
+\left\{  \frac{\rho^{\alpha} +\rho^{1-\alpha}}{2},Y_0  \right\}
\end{equation}
 for any $x \in \mathbb{R}$, then we have

\begin{eqnarray*}
 0 &\le& Tr\left[ {M^* M} \right] \\ 
  &=& \left( {\frac{1}{4}Tr\left[ {\left(i[\rho^{\alpha},X_0]\right)^2  + \left(i[\rho^{1-\alpha},X_0]\right)^2 } \right] + I_{\rho ,\alpha } \left( X \right)} \right)x^2  \\ 
  &&+ \frac{1}{2}Tr\left[ {\left( {i[\rho^{\alpha},X_0] +i[\rho^{1-\alpha},X_0]} \right)\left( {\left\{\rho^{\alpha},Y_0\right\} + \left\{\rho^{1-\alpha},Y_0\right\}} \right)} \right]x \\ 
  &&+ \left( {\frac{1}{4}Tr\left[ {\left\{\rho^{\alpha},Y_0\right\}^2  + \left\{\rho^{1-\alpha},Y_0\right\}^2 } \right] + J_{\rho ,\alpha } \left( Y \right)} \right). \\ 
 \end{eqnarray*}
Therefore we have
\begin{eqnarray*}
 &&\frac{1}{4}\left| {Tr\left[ {\left( {\rho ^\alpha   + \rho ^{1 - \alpha } } \right)^2 \left( {i\left[ {X,Y} \right]} \right)} \right]} \right|^2  \\ 
  &&\hspace*{-8mm}  \le 4\left( {\frac{1}{4}Tr\left[ {\left(i[\rho^{\alpha},X_0]\right)^2  + \left(i[\rho^{1-\alpha},X_0]\right)^2 } \right]
 + I_{\rho ,\alpha } \left( X \right)} \right)\\
&& \times \left( {\frac{1}{4}Tr\left[ {\left\{\rho^{\alpha},Y_0\right\}^2  + \left\{\rho^{1-\alpha},Y_0\right\}^2 } \right] + J_{\rho ,\alpha } \left( Y \right)} \right), \\ 
 \end{eqnarray*}
since we have 
$$Tr\left[ {\left( {i[\rho^{\alpha},X_0] + i[\rho^{1-\alpha},X_0]} \right)\left( {\left\{\rho^{\alpha},Y_0\right\} + \left\{\rho^{1-\alpha},Y_0\right\}} \right)} \right] = Tr\left[ {\left( {\rho ^\alpha   + \rho ^{1 - \alpha } } \right)^2 \left( {i\left[ {X,Y} \right]} \right)} \right].$$
As similar as we have
\begin{eqnarray*}
 &&\frac{1}{4}\left| {Tr\left[ {\left( {\rho ^\alpha   + \rho ^{1 - \alpha } } \right)^2 \left( {i\left[ {X,Y} \right]} \right)} \right]} \right|^2  \\ 
  &&\hspace*{-8mm} \le 4\left( {\frac{1}{4}Tr\left[ {\left(i[\rho^{\alpha},Y_0]\right)^2  + \left(i[\rho^{1-\alpha},Y_0]\right)^2 } \right] 
+ I_{\rho ,\alpha } \left( Y \right)} \right) \\
&& \times \left( {\frac{1}{4}Tr\left[ {\left\{\rho^{\alpha},X_0\right\}^2  + \left\{\rho^{1-\alpha},X_0\right\}^2 } \right] 
+ J_{\rho ,\alpha } \left( X \right)} \right). \\ 
 \end{eqnarray*}
By the above two inequalities, we have
\[
W_{\rho, \alpha}  \left( X \right) W_{\rho,\alpha}  \left( Y \right)
 \geq \frac{1}{4}\left| {Tr\left[ {\left( {\frac{{\rho ^\alpha   + \rho ^{1 - \alpha } }}{2}} \right)^2 {\left[ {X,Y} \right]} } \right]} \right|^2. 
\]

\hfill
\qed

\begin{Cor}
For a quantum state $\rho$ and observables (possibly unbounded operators) $X,Y$ and $\alpha\in[0,1]$, if we have the relation $ [X,Y]=\frac{1}{2\pi i}I$ on
$\mathbf{dom}(XY) \cap \mathbf{dom}(YX)$ and $\rho$ is expressed by
 $\rho = \sum_k \lambda_k \vert \phi_k \rangle \langle \phi_k\vert$, $\vert \phi_k \rangle \in \mathbf{dom}(XY)  \cap \mathbf{dom}(YX)$, then
\[
W_{\rho,\alpha}(X)W_{\rho,\alpha}(Y) \geq  \frac{1}{4}\left| {Tr\left[ {\rho \left[ {X,Y} \right]} \right]} \right|^2.
\]
\end{Cor}
{\it Proof}:
It follows from Theorem \ref{the2_1} and the following inequality: 
$$\frac{1}{4}\left| {Tr\left[ {\left( {\frac{{\rho ^\alpha   + \rho ^{1 - \alpha } }}{2}} \right)^2 \left[ {X,Y} \right]} \right]} \right|^2
\geq \frac{1}{4}\left| {Tr\left[ {\rho \left[ {X,Y} \right]} \right]} \right|^2,$$ whenever we have the canonical commutation relation such as $[X,Y]=\frac{1}{2\pi i}I$.  

\hfill
\qed

\begin{remark}  \label{remarks}
Theorem \ref{the2_1} is not trivial one in the sense of the following (i) and (ii).
\begin{itemize}
\item[(i)] 
Since the arithmetic mean is greater than the geometric mean,
$Tr\left[\left(i \left[\rho^{\alpha},X_0\right]\right)^2\right] \geq 0$ and $Tr\left[\left(i \left[\rho^{1-\alpha},X_0\right]\right)^2\right] \geq 0$ 
imply $ K_{\rho ,\alpha } \left( X \right) \ge I_{\rho ,\alpha } \left( X \right)$, by the use of Schwarz's inequality.
Similarly, $Tr\left[\left\{\rho^{\alpha},Y_0\right\}^2\right] \geq 0$ and $Tr\left[ \left\{\rho^{1-\alpha},Y_0\right\}^2\right] \geq 0$ 
imply $L_{\rho ,\alpha } \left( Y \right) \ge J_{\rho ,\alpha } \left( Y \right)$.  
We then have $W_{\rho,\alpha}  \left( X \right) \ge U_{\rho,\alpha}  \left( X \right)$.

From the inequality (\ref{note2}) and the above, our situation is that we have
$$
U_{\rho,\alpha}(H) \leq U_{\rho}(H)
$$
and
$$
U_{\rho,\alpha}(H) \leq W_{\rho,\alpha}(H).
$$

Our third concern is the ordering between $U_{\rho}(H)$ and  $W_{\rho,\alpha}(H)$.
However, we have no ordering between them. Because we have the follwoing examples.
We set \[
\rho  = \left( \begin{array}{l}
 0.8\,\,\,\,\,\,\,0.0 \\ 
 0.0\,\,\,\,\,\,\,0.2 \\ 
 \end{array} \right),H = \left( \begin{array}{l}
 2.0\,\,\,\,\,3.0 \\ 
 3.0\,\,\,\,\,1.0 \\ 
 \end{array} \right).
\]
If we take $\alpha=0.8$, then $U_{\rho}(H) -  W_{\rho,\alpha}(H)$ approximately takes $-0.0241367$.
If we take $\alpha=0.9$, then $U_{\rho}(H) -  W_{\rho,\alpha}(H)$ approximately takes $0.404141$. 
This example actually shows that there exists a triplet of $\alpha$, $\rho$ and $H$ 
such that $W_{\rho,\alpha}(H) < V_{\rho}(H)$, since we have $U_{\rho}(H) \leq V_{\rho}(H)$ in general.

\item[(ii)] We have no ordering between $
\left| {Tr\left[ {\left( {\frac{{\rho ^\alpha   + \rho ^{1 - \alpha } }}{2}} \right)^2 {\left[ {X,Y} \right]} } \right]} \right|^2 
$ and $
\left| {Tr\left[ {\rho {\left[ {X,Y} \right]} } \right]} \right|^2$,
by the follwoing examples.
If we take 
\[
\rho  = \frac{1}{7}\left( \begin{array}{l}
 \,\,\,\,\,2\,\,\,\,\,\,\,\,\,2i\,\,\,\,\,\,\,\,\,\,1 \\ 
 - 2i\,\,\,\,\,\,\,\,3\,\,\,\, -2i \\ 
 \,\,\,\,\,1\,\,\,\,\,\,\,\,\,\,2i\,\,\,\,\,\,\,\,\,2 \\ 
 \end{array} \right),X = \left( \begin{array}{l}
 \,3\,\,\,\,\,\,\,\,\,3\,\,\,\,\,\,\, -i \\ 
 \,3\,\,\,\,\,\,\,\,\,1\,\,\,\,\,\,\,\,\,\,\,0 \\ 
 \,i\,\,\,\,\,\,\,\,\,\,0\,\,\,\,\,\,\,\,\,\,\,1 \\ 
 \end{array} \right),Y = \left( \begin{array}{l}
 \,\,\,\,1\,\,\,\,\,\,\,\,\,\, - i\,\,\,\,\,\,\,1 - i \\ 
 \,\,\,\,i\,\,\,\,\,\,\,\,\,\,\,\,\,\,\,\,\,1\,\,\,\,\,\,\,\,\,\,\,\,\,\,i \\ 
 1 + i\,\,\,\,\,\, - i\,\,\,\,\,\,\,\,\,\,\,\,3 \\ 
 \end{array} \right),
\]
then we have 
\[
\left| {Tr\left[ {\left( {\frac{{\rho ^\alpha   + \rho ^{1 - \alpha } }}{2}} \right)^2 \left[ {X,Y} \right]} \right]} \right|^2  \simeq 0.348097,\left| {Tr\left[ {\rho \left[ {X,Y} \right]} \right]} \right|^2  \simeq 0.326531.
\]
If we take
\[
\rho  = \frac{1}{7}\left( \begin{array}{l}
 \,\,\,\,\,2\,\,\,\,\,\,\,\,\,2i\,\,\,\,\,\,\,\,\,\,1 \\ 
 - 2i\,\,\,\,\,\,\,\,3\,\,\,\, -2i \\ 
 \,\,\,\,\,1\,\,\,\,\,\,\,\,\,\,2i\,\,\,\,\,\,\,\,\,2 \\ 
 \end{array} \right),X = \left( \begin{array}{l}
 \,3\,\,\,\,\,\,\,\,\,3\,\,\,\,\,\,\, -i \\ 
 \,3\,\,\,\,\,\,\,\,\,1\,\,\,\,\,\,\,\,\,\,\,0 \\ 
 \,i\,\,\,\,\,\,\,\,\,\,0\,\,\,\,\,\,\,\,\,\,\,1 \\ 
 \end{array} \right),Y = \left( \begin{array}{l}
 \,\,\,1\,\,\,\,\,\,\, - i\,\,\,\,\,\,\,\,\,\,0 \\ 
 \,\,\,i\,\,\,\,\,\,\,\,\,\,\,\,\,1\,\,\,\,\,\,\,\,\,\,\,\,i \\ 
 \,\,\,0\,\,\,\,\,\,\, - i\,\,\,\,\,\,\,\,\,\,\,3 \\ 
 \end{array} \right),
\]
then we have
\[
\left| {Tr\left[ {\left( {\frac{{\rho ^\alpha   + \rho ^{1 - \alpha } }}{2}} \right)^2 \left[ {X,Y} \right]} \right]} \right|^2  \simeq 0.304377,\left| {Tr\left[ {\rho \left[ {X,Y} \right]} \right]} \right|^2  \simeq 0.326531.
\]
\end{itemize}
\end{remark}

\begin{remark}
\begin{itemize}
\item[(i)] If we take $M = \rho^{1/2}X_0 x + \rho^{1/2}Y_0$ for any $x \in \mathbb{R}$ presented in Eq.(\ref{def_K}), 
we recover the Heisenberg uncertainty relation Eq.(\ref{HUL}) shown in \cite{Hei}.
\item[(ii)] If we take $\alpha =\frac{1}{2}$, then we recover the inequality (\ref{UL_U}) presented in \cite{Luo1}. 
\item[(iii)] We have another inequalities which are different from the inequality (\ref{ineq_the2_1}), by taking different self-adjoint operators $M$ appeared in the
proof of Theorem \ref{the2_1}. 
\end{itemize}
\end{remark}
\begin{Con}
Our fourth concern is whether the following inequality: 
\begin{equation}  \label{conj_02}
U_{\rho,\alpha}(X)U_{\rho,\alpha}(Y)\geq 
\frac{1}{4}\left| {Tr\left[ {\left( {\frac{{\rho ^\alpha   + \rho ^{1 - \alpha } }}{2}} \right)^2 \left[ {X,Y} \right]} \right]} \right|^2 
\end{equation}
holds or not. However we have not found its proof and any counter-examples yet.
\end{Con}

$K_{\rho,\alpha}(H)$ and $L_{\rho,\alpha}(H)$ are respectively rewritten by
$$
K_{\rho,\alpha}(H) = Tr\left[ \left(\frac{\rho^{\alpha}+\rho^{1-\alpha}}{2}\right)^2H_0^2
 - \left(\frac{\rho^{\alpha}+\rho^{1-\alpha}}{2}\right)H_0\left(\frac{\rho^{\alpha}+\rho^{1-\alpha}}{2}\right)H_0  \right]
$$
and
$$
L_{\rho,\alpha}(H) = Tr\left[ \left(\frac{\rho^{\alpha}+\rho^{1-\alpha}}{2}\right)^2H_0^2
 + \left(\frac{\rho^{\alpha}+\rho^{1-\alpha}}{2}\right)H_0\left(\frac{\rho^{\alpha}+\rho^{1-\alpha}}{2}\right)H_0  \right].
$$
Thus we have 
$$
\frac{1}{2} Tr\left[ \left(i \left[ \frac{\rho^{\alpha}+\rho^{1-\alpha}}{2},H_0 \right]\right)^2\right]
=
\frac{1}{2} Tr\left[ \left(i \left[ \frac{\rho^{\alpha}+\rho^{1-\alpha}}{2},H \right]\right)^2\right]
$$
but we have
$$
\frac{1}{2} Tr\left[ \left( \left\{ \frac{\rho^{\alpha}+\rho^{1-\alpha}}{2},H_0 \right\}\right)^2\right]
\neq
\frac{1}{2} Tr\left[ \left( \left\{ \frac{\rho^{\alpha}+\rho^{1-\alpha}}{2},H \right\}\right)^2\right].
$$
In addition, we have $L_{\rho,\alpha}(H) \geq K_{\rho,\alpha}(H)$ which implies
$$
W_{\rho,\alpha}(H) \equiv  \sqrt{K_{\rho,\alpha}(H)L_{\rho,\alpha}(H)} \geq  \sqrt{K_{\rho,\alpha}(H)K_{\rho,\alpha}(H)}  \geq K_{\rho,\alpha}(H).
$$
Therefore our fifth concern is whether the following inequality for $\alpha \in [0,1]$ holds or not: 

\begin{equation}  \label{conj_03}
K_{\rho,\alpha}(X)K_{\rho,\alpha}(Y)\geq 
\frac{1}{4}\left| {Tr\left[ {\left( {\frac{{\rho ^\alpha   + \rho ^{1 - \alpha } }}{2}} \right)^2 \left[ {X,Y} \right]} \right]} \right|^2. 
\end{equation}
However this inequality fails, because we have a counter-example.
If we set $\alpha = \frac{1}{2}$ and
 \[
\rho  = \frac{1}{4}\left( \begin{array}{l}
 3\,\,\,\,\,0 \\ 
 0\,\,\,\,\,1 \\ 
 \end{array} \right),X = \left( \begin{array}{l}
 \,\,0\,\,\,\,\,\,\,\,i \\ 
  - i\,\,\,\,\,\,\,0 \\ 
 \end{array} \right),Y = \left( \begin{array}{l}
 0\,\,\,\,\,1 \\ 
 1\,\,\,\,\,0 \\ 
 \end{array} \right).
\]
Then we have,
$$K_{\rho,\alpha}(X) K_{\rho,\alpha}(Y) = I_{\rho}(X) I_{\rho}(Y) = \left(\frac{1-\sqrt{3}}{2}\right)^2$$
and
$$\frac{1}{4}\left| {Tr\left[ {\left( {\frac{{\rho ^\alpha   + \rho ^{1 - \alpha } }}{2}} \right)^2 \left[ {X,Y} \right]} \right]} \right|^2
= \frac{1}{4}\left| Tr\left[  \rho \left[ X,Y \right] \right] \right|^2 = \frac{1}{4}. $$
Thus the inequality (\ref{conj_03}) does not hold in general.

Before closing this section, we reconsider the ordering $W_{\rho,\alpha}(H)$ and $V_{\rho}(H)$, although we have already stated 
an example of the triplet $\alpha, \rho$ and $H$ satsfying 
$W_{\rho,\alpha}(H) < V_{\rho}(H)$  in the last line of (i) of 
Remark \ref{remarks}.
If we set $\alpha = \frac{1}{5}$ and
 \[
\rho  = \left( \begin{array}{l}
 \,0.3\,\,\,\,\,0.45 \\ 
 0.45\,\,\,\,\,0.7 \\ 
 \end{array} \right),H = \left( \begin{array}{l}
 1\,\,\,\,\,\,\,3 \\ 
 3\,\,\,\,\,\,\,1 \\ 
 \end{array} \right).
\]
Then $V_{\rho}(H) - W_{\rho,\alpha}(H)$ approximately takes $-0.3072$.
If we set $\alpha = \frac{1}{5}$ and
 \[
\rho  = \left( \begin{array}{l}
 0.3\,\,\,\,\,0.4 \\ 
 0.4\,\,\,\,\,0.7 \\ 
 \end{array} \right),H = \left( \begin{array}{l}
 1\,\,\,\,\,\,\,3 \\ 
 3\,\,\,\,\,\,\,1 \\ 
 \end{array} \right).
\]
Then $V_{\rho}(H) - W_{\rho,\alpha}(H)$ approximately takes $0.682011$.
Therefore we have no ordering between $W_{\rho,\alpha}(H)$ and $V_{\rho}(H)$.
Thus it is natural for us to have an interest in the following conjecture,
 since we have $K_{\rho,\alpha}(H) \leq W_{\rho,\alpha}(H)$ in general.
\begin{Con}
Our final concern is whether the following inequality: 
\begin{equation}  \label{conj_fin}
K_{\rho,\alpha}(H) \leq V_{\rho}(H),\,\,\,\alpha\in[0,1]
\end{equation}
holds or not. However we have not found its proof and any counter-examples yet.
\end{Con}
\section{Concluding remarks}
As we have seen, we introduced a generalized Wigner-Yanase skew information $K_{\rho,\alpha}(H)$
 and then defined a new quantity $W_{\rho,\alpha}(H)$.
We note that our generalied Wigner-Yanase skew information $K_{\rho,\alpha}(H)$ is different type of the Wigner-Yanase-Dyson skew information $I_{\rho,\alpha}(H)$.
For the quantity $K_{\rho,\alpha}(H)$, we do not have a trace inequality related to an uncertainty relation. 
However, we showed that we have a trace inequality related to an uncertainty relation for the quantity $W_{\rho,\alpha}(H)$.
This inequality is a non-trivial one-parameter extension of the uncertainty relation Eq.(\ref{UL_U}) shown by S.Luo in \cite{Luo1}.
In addition, we studied several trace inequaities on informational quantities.

Finally, we give another generalized trace inequality of the inequality (\ref{UL_U}).
For a quantum state $\rho$ an observable $H$ and $\alpha \in [0,1]$, we define
$$
Z_{\rho,\alpha}(H) \equiv \frac{1}{4}\sqrt{Tr\left[(i[\rho^{\alpha},H_0])^2 \right]Tr\left[(i[\rho^{1-\alpha},H_0])^2 \right]Tr\left[\{ \rho^{\alpha},H_0 \}^2 \right]Tr\left[\{ \rho^{1-\alpha},H_0 \}^2 \right]},
$$
with $H_0 \equiv H- Tr[\rho H]I$.
Then we have the following inequality
\begin{equation}  \label{another_generalization}
\sqrt{Z_{\rho,\alpha}(X)Z_{\rho,\alpha}(Y)} \geq 
\frac{1}{4}\left|Tr\left[\rho^{2\alpha}[X,Y] \right]Tr\left[\rho^{2(1-\alpha)}[X,Y] \right] \right|,  
\end{equation}
for a quantum state $\rho$, two observables $X,Y$ and $\alpha \in [0,1]$. 
We note that the inequality (\ref{another_generalization}) recovers the inequality (\ref{UL_U})
by taking $\alpha = 1/2$ and
we do not have any weak-strong relation between 
the inequality (\ref{ineq_the2_1}) and the inequality (\ref{another_generalization}).  
 
\bibliographystyle{amsplain}

\end{document}